# Multicascade-linked synthetic wavelength digital holography using an optical-comb-referenced frequency synthesizer

Masatomo Yamagiwa[1,2], Takeo Minamikawa[1,2], Clément Trovato[3,4], Takayuki Ogawa[2,3], Dahi Ghareab Abdelsalam Ibrahim[1,2,5], Yusuke Kawahito[6], Ryo Oe[2,3], Kyuki Shibuya[2,3], Takahiko Mizuno[1,2], Emmanuel Abraham[7], Yasuhiro Mizutani[2,8], Tetsuo Iwata[1,2], Hirotsugu Yamamoto[2,9], Kaoru Minoshima[2,10], and Takeshi Yasui[1,2]

[1]*Graduate School of Technology, Industrial and Social Sciences, Tokushima University, 2-1, Minami-Josanjima, Tokushima, Tokushima 770-8506, Japan*

[2]*JST, ERATO, MINOSHIMA Intelligent Optical Synthesizer Project, 2-1, Minami-Josanjima, Tokushima, Tokushima 770-8506, Japan*

[3]*Graduate School of Advanced Technology and Science, Tokushima University, 2-1, Minami-Josanjima, Tokushima, Tokushima 770-8506, Japan*

[4]*College of Sciences and Technology, University of Bordeaux, 351 cours de la Libération, Talence Cedex 33405, France*

[5]*Engineering and Surface Metrology Laboratory, National Institute of Standards, Tersa St., El haram, El Giza, Egypt*

[6]*Faculty of Science and Technology, Tokushima University, 2-1, Minami-Josanjima, Tokushima, Tokushima 770-8506, Japan*

[7]*Laboratoire Ondes et Matie`re d'Aquitaine, CNRS UMR 5798, Bordeaux University, Bordeaux Cedex 33000, France*

[8]*Graduate School of Engineering, Osaka University, 2-1, Yamadaoka, Suita, Osaka 565-0871, Japan*

[9]*Center for Optical Research and Education, Utsunomiya University, 7-1-2, Yoto, Utsunomiya, Tochigi 321-8585, Japan*

[10]*Graduate School of Informatics and Engineering, The University of Electro-Communications, 1-5-1 Chofugaoka, Chofu, Tokyo 182-8585, Japan*

Abstract

Digital holography (DH) is a promising method for non-contact surface topography because the reconstructed phase image can visualize the nanometer unevenness in a sample. However, the axial range of this method is limited to the range of the optical wavelength due to the phase wrapping ambiguity. Although the use of two different wavelengths of light and the resulting synthetic wavelength, i.e., synthetic wavelength DH, can expand the axial range up to a few tens of microns, this method is still insufficient for practical applications. In this article, a tunable external cavity laser diode phase-locked to an optical frequency comb, namely, an optical-comb-referenced frequency synthesizer, is effectively used for multiple synthetic wavelengths within the range of 32 μm to 1.20 m. A multiple cascade link of the phase images among an optical wavelength (= 1.520 μm) and 5 different synthetic wavelengths (= 32.39 μm, 99.98 μm, 400.0 μm, 1003 μm, and 4021 μm) enables the shape measurement of a reflective millimeter-sized stepped surface with the axial resolution of 34 nm. The axial dynamic range, defined as the ratio of the maximum axial range (= 0.60 m) to the axial resolution (= 34 nm), achieves $1.7 \times 10^8$, which is much larger than that of previous synthetic wavelength DH. Such a wide axial dynamic range capability will further expand the application field of DH for large objects with meter dimensions.

# 1. Introduction

Digital holography (DH) [1-4] has attracted attention as a three-dimensional (3D) imaging tool with nanometer axial resolution for biomedical imaging [5,6] and industrial inspection [7-9]. In DH, an interference fringe, formed by interfering diffracted light from an object with a reference light, is acquired using a digital imaging sensor such as a charge-coupled device (CCD) or a complementary metal-oxide semiconductor (CMOS) camera. Then, the amplitude and phase images of the object light can be reconstructed by a diffraction calculation of the acquired interference image with a computer. DH is featured by phase imaging, digital focusing, real-time imaging, and quantitative analysis. In particular, phase imaging enables nanometer axial resolution in the 3D shape measurement of transparent or reflective objects. When DH is performed using single-wavelength continuous wave (CW) laser light, the maximum axial range is limited within a half wavelength ($\lambda/2$) for reflective objects or a full wavelength ($\lambda$) for transparent objects due to the phase wrapping ambiguity. Although phase unwrapping processes can expand the maximum axial range [10], their adoption has been limited to a smooth shaped profile.

To extend the maximum axial range over $\lambda/2$ without the need for a phase unwrapping process, synthetic wavelength DH, referred to as SW-DH, has been proposed [11-15]. In this method, DH is performed at two different wavelengths ($\lambda_1$, $\lambda_2$), and then the synthetic wavelength $\Lambda$ between them [$= (\lambda_1\lambda_2)/|\lambda_2-\lambda_1|$] is used to increase the maximum axial range up to $\Lambda/2$, which is larger than $\lambda/2$. However, since stable CW lasers operate at several discrete wavelengths of 532 nm, 612 nm, 633 nm, or 780 nm, the available $\Lambda$ was limited to several microns. Even though three different wavelength lights were used for SW-DH, the maximum $\Lambda$ value remained at approximately a few tens of microns [12,16]. One possible method to further increase $\Lambda$ is the use of a tunable CW laser [17,18]. Although the synthetic wavelength was extended from 3.2 to 40 mm, its wavelength fluctuation and/or mode hopping behavior hinders us from using two slightly different wavelengths of light for the generation of further longer synthetic wavelengths [18]. Furthermore, the axial precision was remained at 35 μm due to no cascade link between phase images with synthetic wavelengths and an optical wavelength. If multiple synthetic wavelengths with high stability and accuracy can be arbitrarily generated in the micrometer to meter range from a single light source and their phase images are cascade linked with each other, the maximum axial range will be significantly greater than the millimeter scale while maintaining the nanometer-scale axial resolution. It is anticipated that the resulting wide axial dynamic range DH will find many applications in the precise profilometry of large objects.

Widely and finely tunable CW light with high stability and accuracy in wavelength or optical frequency can be obtained by an optical-comb-referenced frequency synthesizer (OFS), which is a tunable external cavity laser diode (ECLD) phase-locked to an optical frequency comb (OFC) [19-21]. The OFC is composed of a series of optical frequency modes regularly spaced by a repetition frequency $f_{rep}$ with a carrier-envelope offset frequency $f_{ceo}$. The OFC can be used as an optical frequency ruler by phase locking both $f_{rep}$ and $f_{ceo}$ to a frequency standard. By further phase locking the ECLD to one optical frequency mode of the OFC, the narrow linewidth, high stability, and high accuracy in the OFC are transferred to the ECLD, enabling the determination of the optical frequency based on the frequency standard. Furthermore, the optical frequency of the OFS can be tuned by switching the OFC mode phase-locked by the ECLD or by changing $f_{rep}$ while maintaining the phase locking of the ECLD to the OFC mode. Such an OFS has been used for high-precision broadband spectroscopy in the near-infrared [19] and terahertz [20,21] regions. However, there have been no attempts to generate multiple synthetic wavelengths within the micrometer to millimeter range in SW-DH.

In this article, we demonstrate multiple synthetic wavelength DH (MSW-DH) using an OFS. The OFS is used for the acquisition of multiple phase images at five different synthetic wavelengths and an optical wavelength. The resulting series of phase images were coherently cascade linked for surface profilometry of a millimeter-stepped structure with nanometer axial resolution.

## 2. Principle of operation

### 2.1 Optical-comb-referenced frequency synthesizer (OFS)

Figure 1 shows the schematic diagram of the OFC and OFS. A vast number of OFC modes (freq. = $\nu_m$) can be used as optical frequency markers in a broad spectral range based on the following equation

$$\nu_m = f_{ceo} + m f_{rep}, \tag{1}$$

where $m$ is the mode number of the OFC. The parameter $\nu_m$ is given with an uncertainty of a frequency standard by determining $m$ with an optical wavemeter while phase locking $f_{ceo}$ and $f_{rep}$ to the frequency standard. Next, we consider that the ECLD is phase-locked to the $m$-th OFC mode, namely, the OFS. The optical frequency of OFS $\nu_{ofs}$ is given by

$$\nu_{ofs} = \nu_m + f_{beat} = f_{ceo} + mf_{rep} + f_{beat}, \qquad (2)$$

where $f_{beat}$ is the beat frequency between the $m$-th OFC mode and ECLD light. If $f_{beat}$ is phase-locked to the frequency standard, then the frequency uncertainty in the OFC is transferred to the ECLD, while the ECLD maintains its inherent characteristics such as wide wavelength tunability and moderate power usage. In other words, the OFS is traceable to the frequency standard via the OFC. More importantly from the viewpoint of MSW-DH, $\nu_{ofs}$ can be discretely tuned at a step of $f_{rep}$ within the whole tunable range of the ECLD or the spectral range of the OFC by switching the $m$ value in Eq. (2). Furthermore, $f_{ofs}$ can be tuned more precisely or more continuously by changing $f_{rep}$ while maintaining the phase locking of the ECLD to the OFC mode. Here, we use the $f_{rep}$-step discrete tuning of the OFS for the generation of multiple synthetic wavelengths with a wide dynamic range.

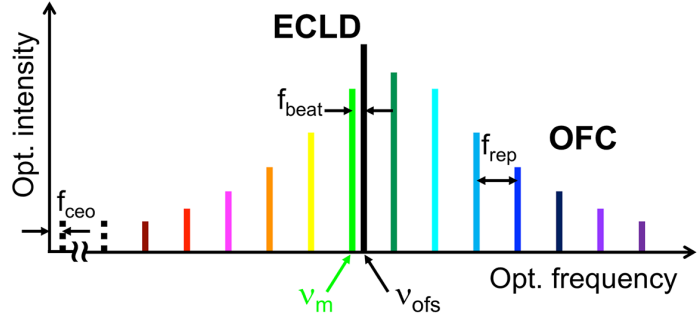

Fig. 1. Optical frequency comb and optical-comb-referenced frequency synthesizer.

*2.2 Multicascade link of the synthetic wavelengths and optical wavelength*

Principle of operation in SW-DH using two different wavelengths ($\lambda_1$, $\lambda_2$) is given in detail elsewhere [11-15]. The synthetic wavelength $\Lambda$ is given by

$$\Lambda = \frac{\lambda_1 \lambda_2}{|\lambda_2 - \lambda_1|}, \qquad (3)$$

In the phase-image-based shape measurement in a reflection configuration, the spatial height distribution $H(x, y)$ of a sample is given by

$$H(x,y) = \left[ N_{\lambda_1}(x,y) + \frac{\phi_{\lambda_1}(x,y)}{2\pi} \right] \frac{\lambda_1}{2} = \left[ N_{\lambda_2}(x,y) + \frac{\phi_{\lambda_2}(x,y)}{2\pi} \right] \frac{\lambda_2}{2}$$
$$= \left[ \frac{\phi_\Lambda(x,y)}{2\pi} \right] \frac{\Lambda}{2}, \qquad (4)$$

where $N_\lambda(x, y)$ and $\phi_\lambda(x, y)$ are the spatial distributions of interference-fringe orders (integer) and phase values at $\lambda_1$ or $\lambda_2$ and $N_\Lambda(x, y)$ and $\phi_\Lambda(x, y)$ are those at $\Lambda$, respectively. The function $\phi_\Lambda(x, y)$ is calculated by taking the difference of $\phi_\lambda(x, y)$ between $\lambda_1$ and $\lambda_2$. We here assumed that phase wrapping did not occur in $\phi_\Lambda(x, y)$. While $\phi_\lambda(x, y)$ and $\phi_\Lambda(x, y)$ can be measured within the phase range of 0 to $2\pi$ rad, $N_\lambda(x, y)$ cannot be directly determined due to the phase wrapping ambiguity. If $\phi_\Lambda(x, y)$ is used to determine $N_\lambda(x, y)$ and later $H(x,y)$ by the cascade link between $\lambda$ and $\Lambda$, the maximum axial range is expanded to $\Lambda/2$ (typically, ~ $10\lambda$) whereas the axial resolution of approximately $\lambda/100$ ~ $\lambda/1000$ is maintained. The resulting axial dynamic range is expanded to $10^3$ ~ $10^4$. However, if $\phi_\Lambda(x, y)$ also undergoes phase wrapping, then $H(x,y)$ cannot be determined. The limited synthetic wavelengths (typically < a few tens of microns) by available CW lasers hinder the increase in the axial dynamic range.

The OFS can generate multiple synthetic wavelengths with different orders of magnitude (= $\Lambda_1 < \Lambda_2 < \cdots < \Lambda_{n-1} < \Lambda_n$) as multiple cascades, as shown in Fig. 2. When such a series of multiple synthetic wavelengths were used for MSW-DH together with $\lambda$, $H(x, y)$ is given by

$$H(x,y) = \left[ N_\lambda(x,y) + \frac{\phi_\lambda(x,y)}{2\pi} \right] \frac{\lambda}{2} = \left[ N_{\Lambda_1}(x,y) + \frac{\phi_{\Lambda_1}(x,y)}{2\pi} \right] \frac{\Lambda_1}{2}$$
$$= \left[ N_{\Lambda_2}(x,y) + \frac{\phi_{\Lambda_2}(x,y)}{2\pi} \right] \frac{\Lambda_2}{2} = \cdots\cdots$$
$$= \left[ N_{\Lambda_{n-1}}(x,y) + \frac{\phi_{\Lambda_{n-1}}(x,y)}{2\pi} \right] \frac{\Lambda_{n-1}}{2} = \left[ \frac{\phi_{\Lambda_n}(x,y)}{2\pi} \right] \frac{\Lambda_n}{2}, \qquad (5)$$

where $N_{\Lambda i}(x, y)$ and $\phi_{\Lambda i}(x, y)$ are the spatial distributions of interference-fringe orders and phase values at $\Lambda_i$, respectively. If each $N_{\Lambda i}(x, y)$ value can be determined one after another from the longest synthetic wavelength $\Lambda_n$ to the shortest synthetic wavelength $\Lambda_1$, then one can finally determine $N_\lambda(x, y)$ without errors. For example, in Eq. (5), the no-wrapping phase image $\phi_{\Lambda n}(x, y)$ obtained at the longest synthetic wavelength $\Lambda_n$ is used to calculate $H_{\Lambda n}(x, y)$, where $H_{\Lambda n}(x, y)$ is $H(x, y)$ determined by $\Lambda_n$. Then, $H_{\Lambda n}(x, y)$ is used to determine $N_{\Lambda n-1}(x, y)$. Subsequently, the determined $N_{\Lambda n-1}(x, y)$ and the measured $\phi_{\Lambda n-1}(x, y)$ are used to determine $H_{\Lambda n-1}(x, y)$ more precisely. $N_{\Lambda i}(x, y)$ is given by

$$N_{\Lambda_i}(x,y) = INT\left[\frac{H_{\Lambda_{i+1}}(x,y)}{\Lambda_i/2} - \frac{\phi_{\Lambda_i}(x,y)}{2\pi}\right]. \tag{6}$$

By repeating a similar procedure from the longest to the shortest synthetic wavelength to the optical wavelength, $N_\lambda(x, y)$ can be determined correctly. By using such a multicascade link from $\Lambda_n$ to $\lambda$, both the maximum axial range of $\Lambda_n/2$ and the axial resolution of $\lambda/100 \sim \lambda/1000$ can be achieved at the same time. The resulting axial dynamic range becomes several orders of magnitude larger than the previous single SW-DH.

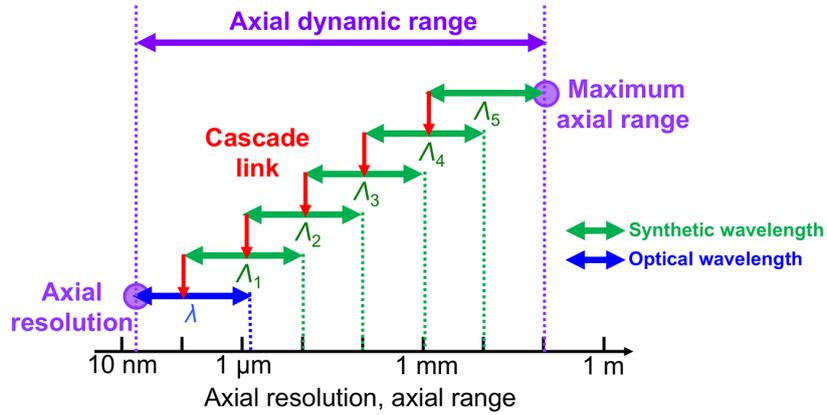

Fig. 2. Principle of operation in MSW-DH.

## 3. Methods

### *3.1 Experimental setup of the OFS*

Figure 3(a) shows an experimental setup of the OFS, which is composed of an OFC and a tunable ECLD. We used a fiber OFC system (OFC; FC1500-250-WG, Menlo Systems GmbH, Martinsried, Germany, center wavelength = 1550 nm, spectral range = 1500 ~ 1600 nm, mean power = 60 mW, $f_{rep}$ = 250 MHz, $f_{ceo}$ = 20 MHz) for a frequency reference of a tunable ECLD (ECLD; LT-5001N, OptoComb, Inc., Tokyo, Japan, tuning range = 1520 nm ~ 1595 nm, mean power = 30 mW). The parameters $f_{rep}$ and $f_{ceo}$ were phase-locked to a rubidium frequency standard (Rb-FS; FS725, Stanford Research Systems, Sunnyvale, CA, USA, accuracy = 5×10$^{-11}$ and instability = 2×10$^{-11}$ at 1 s) by a frequency control system accompanied with the OFC. After polarization adjustment with pairs of a quarter waveplate (Q) and a half waveplate (H), a horizontally polarized beam from the OFC and a vertically polarized beam from the ECLD were spatially overlapped by a polarization beam splitter (PBS1), the polarization of both beams was directed at ± 45 degrees by another H, and their horizontal polarization components were extracted by another PBS (PBS2) for interference. After diffraction at a diffraction grating (GR25-0616, Thorlabs Inc., Newton, NJ, USA, 600 grooves/mm), an optical beat signal with $f_{beat}$ between the ECLD beam and its most adjacent OFC mode was detected by an InGaAs photodetector (PD; PDA10CF-FC, Thorlab, bandwidth = 150 MHz). A proportional-integral servo controller (PI-SC; LB1005, Newport Corp., Irvine, CA, USA, bandwidth = 10 MHz) was used to control the current signal and the intracavity piezoelectric transducer of the ECLD so that $f_{beat}$ was stabilized at 30 MHz by comparing $f_{beat}$ with a reference frequency signal (freq. = 30 MHz) synthesized from a function generator (FG; 33210A, Keysight Technologies, Santa Rosa, CA, USA) using the same Rb-FS as an external time base. For tuning $f_{ofs}$, an $m$ value was selected by changing the tilt angle of an internal grating in the ECLD.

### *3.2 Experimental setup of MSW-DH*

We used an off-axis Michelson-type interferometer for MSW-DH, as shown in Fig. 3(b). The beam diameter of the output light from the OFS (mean power = 3 mW) was expanded to 54 mm by a pair of off-axis parabolic mirrors (OA-PM1, off-axis angle = 90°, diameter = 25.4 mm, focal length = 25.4 mm; OA-PM2, off-axis angle = 90°, diameter = 50.8 mm, focal length = 190.5 mm) and was fed into the Michelson interferometer. In the interferometer, the object beam was reflected by a cubic beam splitter (BS, reflection = 50 %, transmittance = 50 %), reflected at a sample, and then passed through the BS. The reference beam passed through the BS, was reflected at a gold mirror (M, surface roughness = 63 nm), and then was reflected by the BS. Both beams were incident onto a cooled infrared CCD camera (Goldeye P-008, Allied Vision Technol. GmbH, Stadtroda, Germany, 320 x 256 pixels, pixel size = 30.0 μm × 30.0 μm, exposure time = 10 ms, digital output resolution = 14 bit) at an off-axis angle of 0.6°. This process resulted in the generation of interference patterns, namely, the digital hologram.

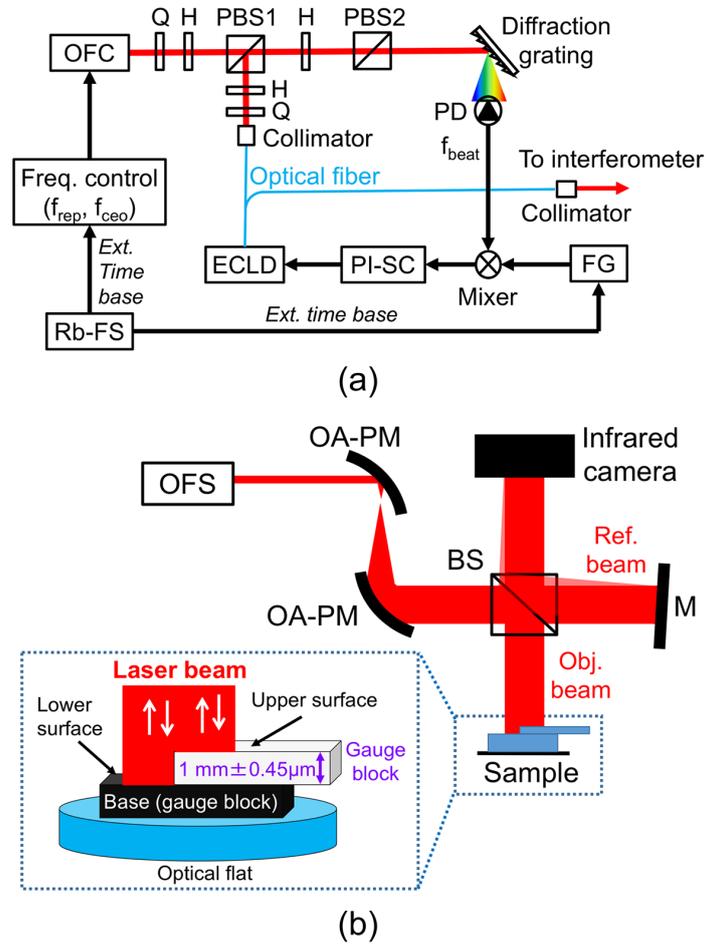

Fig. 3. Experimental setup of the (a) OFS and (b) MSW-DH. OFC: optical frequency comb, ECLD: external cavity laser diode, Rb-FS rubidium frequency standard, PI-SC: proportional-integral servo controller, Hs: 1/2 waveplate, Qs:1/4 waveplate, PBS1 and PBS2: polarization beam splitter, PD: photodetector, OFS: optical-comb-referenced frequency synthesizer, OA-PM1 and OA-PM2: off-axis parabolic mirror, BS: beam splitter, and M: mirror. Inset in Fig. 3(b) shows a schematic drawing of a sample with a stepped surface.

*3.3 Angular spectrum method for the wavefront reconstruction of DH*

We used an angular spectrum method (ASM) [22,23] for the wavefront propagation calculation. The angular spectrum $A(k_x, k_y; z)$ is given by the Fourier transform of the optical field $E(x, y; z)$. Therefore, the angular spectrum at the propagation distance $z = 0$ is given by

$$A_0(k_x, k_y) = F[E_0(x_0, y_0)]$$
$$= \iint E_0(x_0, y_0) \exp[-i(k_x x_0 + k_y y_0)] dx_0 dy_0. \qquad (7)$$

Next, the angular spectrum after propagation is specified as the product of $A_0(k_x, k_y)$ and a phase term for the propagation distance as follows

$$A(k_x, k_y; z) = A_0(k_x, k_y) \exp\left[iz\sqrt{k^2 - k_x^2 - k_y^2}\right]. \qquad (8)$$

Finally, the optical field after propagation is obtained by the inverse Fourier formation of $A(k_x, k_y; z)$ as follows

$$E(x, y; z) = F^{-1}[A(k_x, k_y; z)]$$
$$= F^{-1}\left[F[E_0(x_0, y_0)] \exp\left[iz\sqrt{k^2 - k_x^2 - k_y^2}\right]\right]. \qquad (9)$$

The ASM has three advantages in wavefront reconstruction: First, the digital filtering in the spatial frequency domain eliminates the zero-order diffraction light and the conjugate first-order diffraction light images, both of which are unnecessary for the wavefront reconstruction. The spatial filtering of unnecessary components leads to the improved quality of reconstructed amplitude and phase images of the object. Second, the pixel size of the reconstructed image is the same as that of the obtained digital hologram. Third, the ASM has no limitation for the reconstruction distance $z$ because the optical field is treated as a plane wave.

## 4. Results

*4.1 Basic performance of the OFS*

We first estimated the frequency fluctuation of $\nu_{ofs}$. Since $f_{ceo}$, $f_{rep}$, and $f_{beat}$ in Eq. (2) were phase-locked to the rubidium frequency standard, their frequency instability should be identical to that of the frequency standard. To evaluate the frequency instability of the rubidium frequency standard, we prepared two independent rubidium frequency standards with equivalent performance and used them for the frequency signal generation and the frequency signal measurement with an RF frequency counter (53220A, Keysight Technologies). The resulting frequency instability, calculated by Allan deviation, is shown in Fig. 4(a). From this frequency instability, we estimated the frequency fluctuation of $f_{ceo}$, $f_{rep}$, and $f_{beat}$ with respect to gate time, as shown in Fig. 4(b). We further appended the frequency fluctuation of $mf_{rep}$ when $f_{ofs}$ was fixed at 193.41505 THz or 1549.9955 nm by setting $m$ to be 773,660. A comparison of $f_{ceo}$, $mf_{rep}$, and $f_{beat}$ clearly indicated the frequency fluctuation of $f_{ofs}$ is determined by that of $mf_{rep}$ due to the large number of $m$ values. Importantly, the optical frequency fluctuation of the OFS is less than 1MHz at a gate time of 1 ms, which is several orders of magnitude smaller than that of CW lasers used in the previous research of SW-DH, due to the coherent frequency link with the frequency standard via the OFC.

The present OFS can generate single-mode CW light at a step of $f_{rep}$ (= 250 MHz) or 1.9 pm within the wavelength range of 1520 nm ~ 1595 nm, that can be used for two wavelength lights with a wavelength difference of 1.9 pm to 75 nm. Fig. 4(c) shows a relation between the wavelength difference $\Delta\lambda$ (= $\lambda_2 - \lambda_1$, $\lambda_1$ = 1520 nm) and the corresponding $\Lambda$ [see Eq. (3)], indicating that the available $\Lambda$ ranges from 32 μm to 1.20 m. The maximum $\Lambda$ of 1.20 m is three orders of magnitude larger than that in the previous study [18], highlighting the wide axial dynamic range in MSW-DH. We here generated five different synthetic wavelengths ($\Lambda_1$ = 32.38644 μm, $\Lambda_2$ = 99.97909 μm, $\Lambda_3$ = 400.0234 μm, $\Lambda_4$ = 1,002.524 μm, $\Lambda_5$ = 4021.204 μm) and a single optical wavelength ($\lambda$ = 1.520302 μm) in the following demonstration.

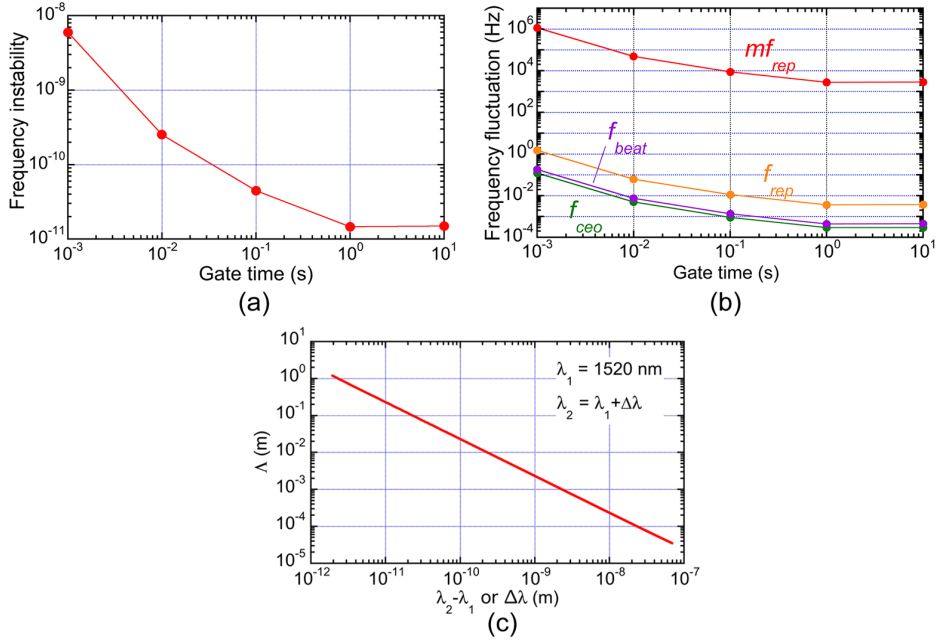

Fig. 4. Performance of the OFS. (a) Frequency instability of a rubidium frequency standard. (b) Frequency fluctuation of $f_{ceo}$, $f_{rep}$, $f_{beat}$, and $mf_{rep}$ in the OFS. (c) Relation of the wavelength difference between two wavelength lights and the synthetic wavelength in the OFS.

*4.2 Evaluation of the phase noise in MSW-DH*

We evaluate the basic performance of the phase imaging by measuring a gauge block (164042, Mitutoyo, Kawasaki, Japan, thickness = 1 mm ± 0.45 μm, surface roughness = 21.6 nm) as a sample. We first evaluated spatial phase noise. We reconstructed the phase images of the sample at a reconstruction distance, $z$, of 134.2 mm using the ASM. Figures 5(a), 5(b), 5(c), 5(d), 5(e), and 5(f) show a series of phase images at $\lambda$, $\Lambda_1$, $\Lambda_2$, $\Lambda_3$, $\Lambda_4$, and $\Lambda_5$, respectively. Each phase image shows the spatial distribution of the phase value at the corresponding wavelength. We here defined the standard deviation of the spatial phase distribution $\phi_\lambda(x, y)$ or $\phi_{\Lambda i}(x, y)$ as a spatial phase noise, limiting the precision of the surface unevenness measurement. Figure 5(g) shows a comparison of the spatial phase noise among the optical wavelength $\lambda$ and synthetic wavelengths $\Lambda_1 \sim \Lambda_5$. Each phase image of $\Lambda_i$ has a phase noise around 0.02 rad, which corresponds to a phase resolving power of around 1/314. The reason for the large phase noise at $\lambda$ is mainly due to the surface roughness of the gauge block or the mirror in the referenced arm. By decreasing the wavelength from $\Lambda_5$ to $\lambda$, the precision of the surface unevenness measurement improved from 6.59 μm to 0.01 μm, as shown in Fig. 5(h).

We next evaluated temporal phase noise. The temporal phase noise was obtained by calculating the standard deviation of the phase values at the same pixel in 100 repetitive phase images with the same wavelength. The temporal phase noise depends on the robustness of the optical systems to an environmental disturbance, such as air turbulence or mechanical vibration, and hence is a critical factor to determine the uncertainty in the height measurement based on the phase image. Figures 6(a), 6(b), 6(c), 6(d), 6(e), and 6(f) show the spatial distribution of the temporal phase noise at $\lambda$, $\Lambda_1$, $\Lambda_2$, $\Lambda_3$, $\Lambda_4$, and $\Lambda_5$, respectively. The similar distribution of temporal phase noise was confirmed in each phase image. Figure 6(g) compares the mean of the temporal phase noise among the optical wavelength $\lambda$ and synthetic wavelengths $\Lambda_1 \sim \Lambda_5$. The temporal phase noise of around 0.1 rad, which corresponds to the phase resolving power of 1/63, respectively, was confirmed at each wavelength; this value is larger than the spatial phase noise due to the phase noise in the time domain. Figure 6(h) shows the corresponding uncertainty of the height measurement. From this result, we can determine the sample height with an axial resolution of 34 nm.

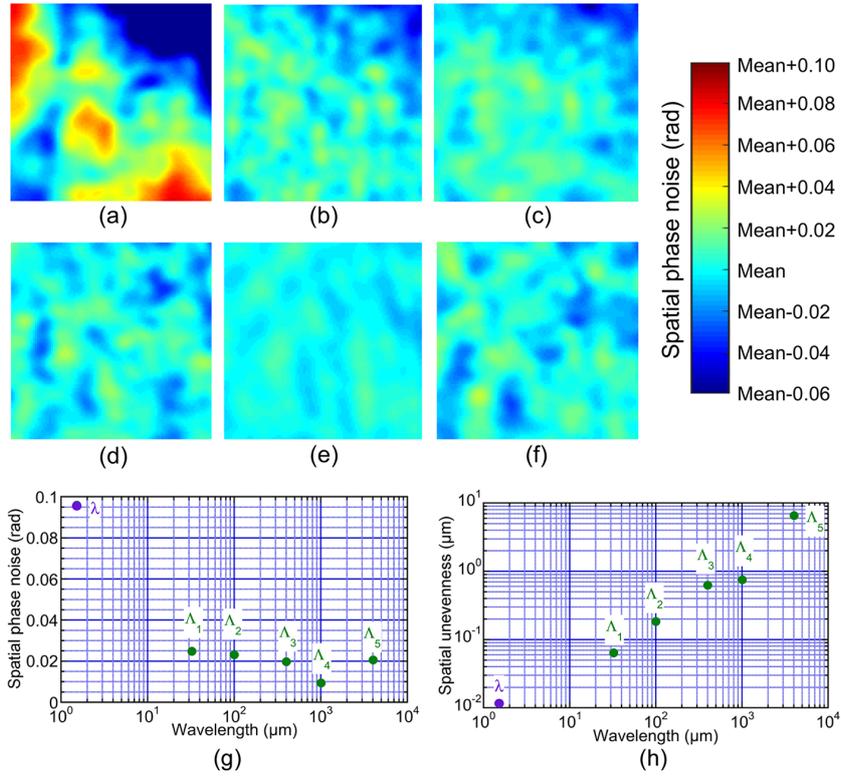

Fig. 5. Spatial phase noise of (a) $\lambda$ (= 1.520302 µm), (b) $\Lambda_1$ (= 32.38644 µm), (c) $\Lambda_2$ (= 99.97909 µm), (d) $\Lambda_3$ (= 400.0234), (e) $\Lambda_4$ (= 1,002.524 µm), and (f) $\Lambda_5$ (= 4021.204 µm). Dependence of (g) spatial phase noise and (h) unevenness precision on wavelength.

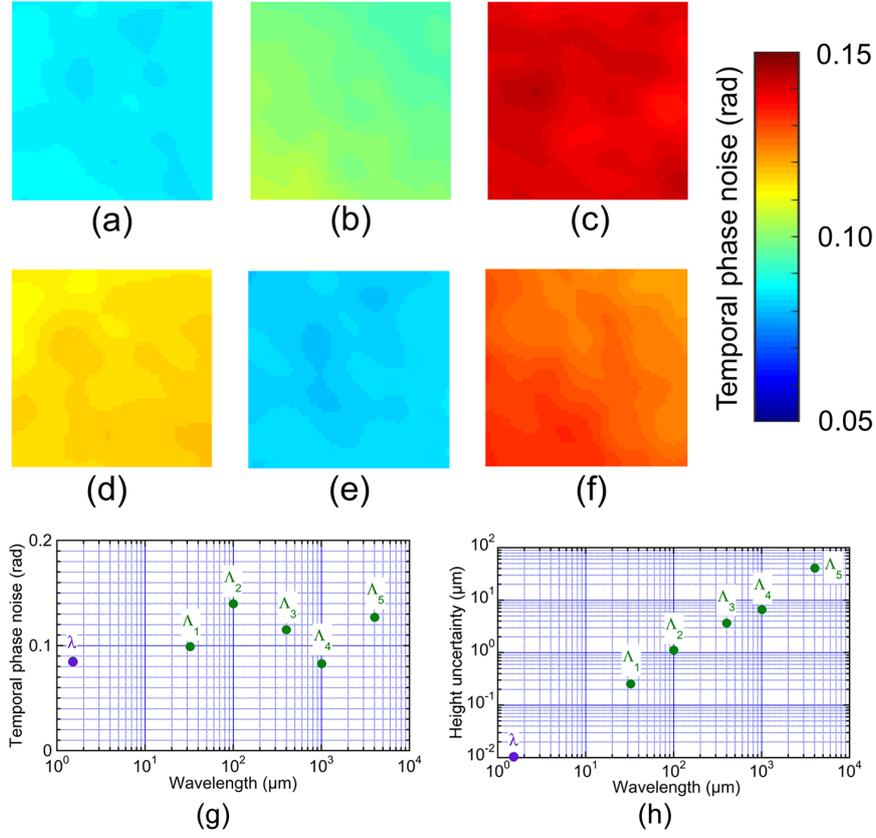

Fig. 6. Temporal phase noise of (a) $\lambda$ (= 1.520302 μm), (b) $\Lambda_1$ (= 32.38644 μm), (c) $\Lambda_2$ (= 99.97909 μm), (d) $\Lambda_3$ (= 400.0234), (e) $\Lambda_4$ (= 1,002.524 μm), and (f) $\Lambda_5$ (= 4021.204 μm). Dependence of (g) temporal phase noise and (h) height uncertainty on wavelength.

*4.3 3D Shape measurement of the stepped surface*

Finally, we performed a 3D shape measurement of a stepped surface based on the cascade-linked phase images. We attached the same gauge block (164042, Mitutoyo, Kawasaki, Japan, thickness = 1 mm ± 0.45 μm, surface roughness = 21.6 nm) to another equivalent gauge block for wringing, as shown in the inset of Fig. 3(b). This stepped surface was used as a sample. Figures 7(a) and 7(b) show the spatial distributions of relative height for the upper surface and lower surface of the sample with respect to the number of cascade links (CLs). For example, no CLs just uses $\Lambda_5$ whereas full CLs is achieved by use of $\Lambda_5$, $\Lambda_4$, $\Lambda_3$, $\Lambda_2$, $\Lambda_1$, and $\lambda$. These results clearly indicated that the precision of the surface unevenness measurement was improved without error of $N_{\Lambda i}(x,$ *y*) or $N_\lambda(x, y)$ by the cascade link among the five different synthetic wavelengths and the optical wavelength.

To evaluate the uncertainty of the step height measurement, we repeated similar experiments for the same sample, and then determined the step height at each number of cascade links. Figure 7(c) shows the mean and standard deviation of the five repetitive measurements with respect to the number of cascade links. Finally, the step height was determined to be 999.969±0.025 μm using the full CLs. Figure 7(d) shows the corresponding 3D shape of the stepped surface. The step difference determined by the MSW-DH was in good agreement with the specification value of the thickness in the gauge block (= 1 mm ± 0.45 μm).

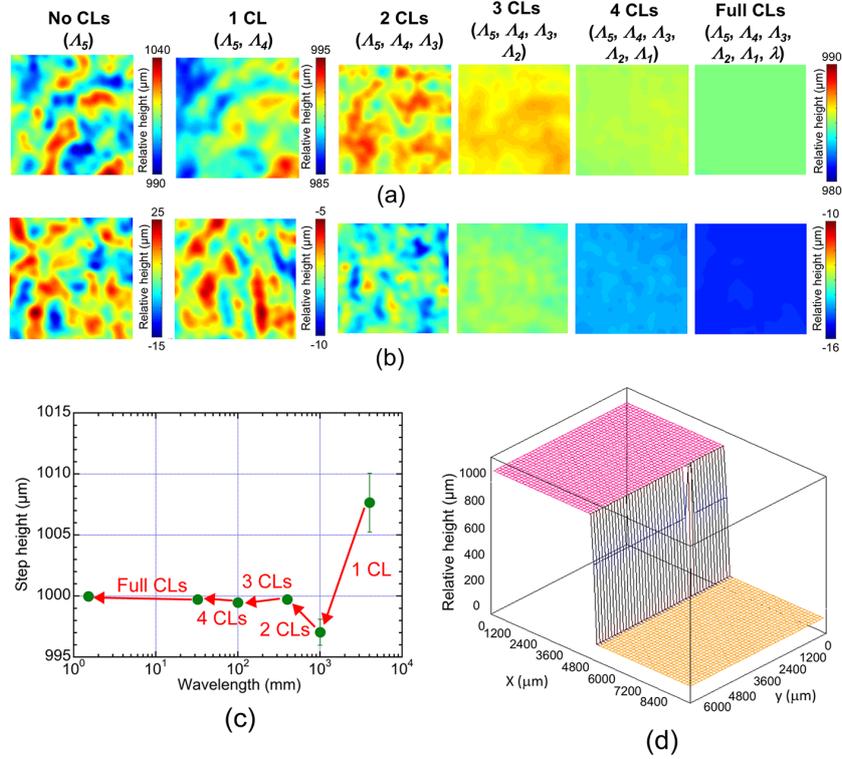

Fig. 7. Surface topography of a stepped surface sample. Spatial distributions of relative height for (a) an upper surface and (b) a lower surface with respect to the number of cascade links (CLs). The image size is 3 mm by 3 mm. (c) Improvement of the precision in the step height measurement with respect to the number of cascade links. (d) 3D profile of a 1-mm-step sample determined by the full cascaded link of a single optical wavelength and 5 synthetic wavelengths.

## 5. Discussion

We first discuss a possibility of a further enhancement in the axial dynamic range in the phase-image-based 3D shape measurement. In the demonstration above, an axial resolution of 34 nm was achieved within the axial range of 2 mm by using cascade link between 5 different synthetic wavelengths and an optical wavelength. However, the present OFS has a potential to increase the axial range up to 0.60 m. The resulting axial dynamic range was $1.7 \times 10^8$. Such a wide axial dynamic range is a distinctive feature in MSW-DH. However, there is still room to further expand the axial range. From the viewpoint of increased synthetic wavelength, the more precise tuning of $f_{ofs}$ can be achieved by changing $f_{rep}$ while maintaining the phase locking of the ECLD to the OFC mode. In this case, the minimum step is limited by the linewidth of the OFS [typically, 10 kHz at a gate time of 0.1 s in Fig. 4(b)]. Two wavelengths of light with an optical frequency difference of 10 kHz results in the generation of several tens kilometers in $\Lambda$, which is four orders of magnitude larger than the present maximum $\Lambda$. On the other hand, from the viewpoint of decreased axial resolution, use of more robust optical system will further reduce the temporal phase noise down to $\lambda/1000$. Such improvement for the increased synthetic wavelength and the decreased axial resolution will further increase the axial dynamic range.

We next discuss the possibility of real-time MSW-DH measurements. In this article, we applied five different synthetic wavelengths and one optical wavelength for MSW-DH. To this end, we acquired holograms at six different optical wavelengths and calculated the phase images at synthetic wavelengths. Those holograms were acquired in order while ECLD was phase-locked to different mode of the OFC; the total acquisition time for the multiple holograms was typically a few minutes. In this article, we performed fine cascade links among phase images with different wavelengths as a proof of concept; however, considering the temporal phase noise at $\Lambda_i$ and $\lambda$, we can further reduce the number of cascade links while maintaining the same performance in MSW-DH, leading to the reduction of the total acquisition time. Furthermore, if the phase locking procedure of ECLD is excluded, the acquisition time of multiple holograms will be largely reduced. One possible method is use of a line-by-line pulse-shaping technique in the OFC [24]. In this case, a single OFC mode is arbitrarily and quickly extracted by the use of a spatial light modulator and then is directly used for the rapid generation of multiple synthetic wavelengths. If the selection of an arbitrary OFC mode is performed in synchronization with a frame timing in the infrared camera, all holograms required for MSW-DH will be acquired in real-time. Such an approach will enable us to achieve real-time MSW-DH.

## 6. Conclusion

We demonstrated wide axial dynamic range MSW-DH using an OFS. The OFS was effectively used for multiple synthetic wavelengths within the range of 32 μm to 1.20 m. The cascade link of phase images was demonstrated among a single optical wavelength and 5 synthetic wavelengths within the range from 1.5 μm to 4000 μm, and enabled us to achieve an axial resolution of 34 nm in a 3D shape measurement of a 1-mm-stepped surface with a precision of 25 nm. The . The wide axial dynamic range MSW-DH can be a powerful tool for industrial inspection of large objects.

## Funding

Exploratory Research for Advanced Technology (ERATO), Japanese Science and Technology Agency (MINOSHIMA Intelligent Optical Synthesizer Project, JPMJER1304); Japan Society for the Promotion of Science (JSPS) (18K04981).

## Acknowledgments

The authors thank Drs. Yasuaki Hori and Akiko Hirai at the National Institute of Advanced Industrial Science and Technology (AIST) for their help in the measurement of the gauge-block sample. We also acknowledge Dr. Yoshiaki Nakajima at the University of Electro-Communications and Dr. Sho Okubo at AIST for their help in the OFS.